\documentclass[aps, prb, reprint, showpacs, amsmath]{revtex4-1}
\usepackage[colorlinks=true, urlcolor=black, citecolor=blue, linkcolor=blue]{hyperref}
\usepackage{graphicx}
\usepackage{bm}
\usepackage{multirow}
\begin{document}

\title{Electronic and mechanical properties of few-layer borophene}
\author{Hongxia Zhong}
\author{Kaixiang Huang}
\author{Guodong Yu}
\author{Shengjun Yuan}
\email{s.yuan@science.ru.nl}
\affiliation{School of Physics and Technology, Wuhan University, Wuhan, 430072,
 People's Republic of China}

\begin{abstract}
We report first principle calculations of electronic and mechanical properties of few-layer borophene with the inclusion of interlayer van der Waals (vdW) interaction. The anisotropic metallic behaviors are preserved from monolayer to few-layer structures. The energy splitting of bilayer borophene at $\Gamma$ point near the Fermi level is about 1.7~eV, much larger than the values (0.5--1~eV) of other layered semiconductors, indicating much stronger vdW interactions in metallic layered borophene. In particular, the critical strains are enhanced by increasing the number of layers, leading to much more flexibility than that of monolayer structure. On the one hand, because of the buckled atomic structures, the out-of-plane negative Poisson's ratios are preserved as the layer-number increases. On the other hand, we find that the in-plane negative Poisson's ratios disappear in layered borophene, which is very different from puckered black phosphorus.  The negative Poisson's ratio will recover if we enlarge the interlayer distance to 6.3~$\mbox\AA$, indicating that the physical origin behind the change of Poisson's ratios is the strong interlayer vdW interactions in layered borophene.
\end{abstract}
\maketitle

\section{Introduction}
Boron, next to carbon in the periodic table, has more than sixteen bulk and numerous low dimensional allotropes\cite{1,2}. The polymorphism is originated from the electron deficiency, resulting in the multi-center B--B bonds, which is much more complicated than that in carbon\cite{2}. Among the polymorphic structures, two-dimensional (2D) boron materials (borophene) have attracted extensive theoretical interests because of their remarkable physical and chemical properties\cite{3,4,5,6}. A class of borophene have therefore been designed\cite{7}. However, there was no evidence that the 2D boron sheets could be fabricated experimentally until 2015\cite{8}. Mannix \emph{et al} first reported the synthesis of 2-P$mmm$ borophene in ultra-high vacuum conditions on Ag (111) surfaces by physical vapor deposition in Ref.~\onlinecite{8}. Scanning tunneling microscopy characterization of the one-atom-thick 2D sheets revealed a hexagonal arrangement of boron atoms with an extra atom in the middle. Later, Feng \emph{et al} have also grown two boron sheets, a $\beta_{12}$ sheet and $\chi_3$ sheet on Ag (111) surfaces, and the $\beta_{12}$ phase is found to be with gapless Dirac cones\cite{9,10,11}. These two phases are in triangular lattices but in flat geometry and with a periodic arrangement of atom vacancies\cite{9,10}. The observed phase depends on the deposition rate and temperature in the experiment, confirming the predicted large polymorphism of borophene.

The successful fabrications of borophene have inspired much follow-up works, especially for the buckled 2-P$mmm$ structure. Owing to its anisotropic atomic structure, the buckled borophene shows highly anisotropic metallic properties, very different from the semi-metallic graphene\cite{12} and semiconducting transition-metal dichalcogenides (TMDCs) \cite{8,13,14,15,16,17,18,19,20,21}. The predicated Fermi velocity of hydrogenated borophene ($3.5\times10^6$~m/s) is nearly four times higher than that of graphene ($8.2\times10^5$~m/s)\cite{22,23,24}. The investigated optical spectra attest high optical transparency (up to 100\% transmission) predicted up to roughly 3 eV, making borophene more transparent than graphene\cite{13}. Borophene thus can be considered as a good candidate of transparent conductive 2D material for photovoltaics and touch screens, due to its robust metallicity, high Fermi velocity, and ultra-high optical transparency. Furthermore, pristine borophene has been predicted to exhibit phonon-mediated superconductivity with critical temperature $T_\mathrm c$ in the range of 10--20~K\cite{25,26}. For the mechanical properties, the buckled borophene shows considerable toughness, and it has been demonstrated that the in-plane Young's modulus along the armchair direction (398~GPa$\cdot$nm)\cite{8} can rival that of graphene (340~GPa$\cdot$nm)\cite{12}. Notably, monolayer borophene is calculated to exhibit negative in-plane and out-of-plane Poisson's ratios. The negative Poisson's ratios of monolayer borophene make it worth further exploitation in many applications, such as medicine\cite{27}, tougher composites\cite{28}, national security and defenses\cite{29,30}. In the following, we focus solely on the buckled 2-P$mmm$ borophene, a promising 2D material which has fantasy physical properties.

Previous theoretical studies of 2-P$mmm$ borophene mainly focused on the physical properties of monolayer borophene\cite{13,14,15}. A comprehensive study of the electronic and elastic properties of the few-layer borophene is still lacking. Moreover, the physical properties of layered 2D materials are highly dependent on their thickness, such as the band crossover in hexagonal TMDCs which is originated from the interlayer van der Waals (vdW) interaction and variations of screening\cite{31,32,33}. Therefore, it is of both fundamental and practical interests to attain a better understanding of the interlayer vdW interactions among the layered borophene. We note that it is inclined to form 3D boron clusters instead of 2D layered borophene by depositing additional boron atoms on monolayer flat $\beta_{12}$ sheet and $\chi_3$ borophene supported by Ag (111) surfaces in Ref.~\onlinecite{9}, as a result of the saturation of the Ag--B interfacial interactions. However, in the experiments there were occasionally small islands observed of second layer, although with different structures comparing to the first layer\cite{9}. We thus believe that few-layer buckled 2-P$mmm$ borophene can be fabricated with different experimental method or a proper choose of the substrate.

In this article, we present a comprehensive theoretical study of the electronic and elastic properties of few-layer borophene, using a first-principle approach including the vdW interaction. We first give a brief description of the numerical methods, and then discuss the chemical bonding nature of borophene, the structure and electronic properties of layered structure with different types of interlayer stacking, and determine the preferable stacking mode which has the lowest total energy. We will then focus on the layered structure with this stacking mode, and investigate the electronic and mechanical properties of bilayer, trilayer, and four-layer borophene. The out-of-plane and in-plane Possion's ratios will be studied in detail, by considering the influence of the interlayer vdW interaction. We will summarize our main findings in the conclusion.

\section{Methods}
Our calculations are performed using the projector augmented wave (PAW) method\cite{34} implemented in the Vienna ab initio simulation package (VASP) code\cite{35,36}. Perdew, Burke, and Ernzerhof (PBE) form of the generalized gradient approximation (GGA) exchange-correlation functional with van der Waals corrections (vdW-DFT)\cite{37,38,39}, and the PAW pseudo potentials\cite{34} are adopted. The cut-off energy is set to 500~eV after convergence tests. A $\Gamma$-centered Monkhorst-Pack $k$-point\cite{40,41} grid of $15\times13\times1$ for one borophene unit cell is chosen for relaxations and the grid of $25\times23\times1$ for property calculations. In our current calculations, the total energy is converged to less than $10^{-5}$~eV. The maximum force is less than 0.02~eV/$\mbox\AA$ during the optimization. A vacuum space between neighboring supercells is set to be more than 25~$\mbox\AA$ to avoid spurious interactions. The Crystal Orbital Hamilton Population (COHP) curves are calculated using the LOBSTE software\cite{42}.

For 2D orthorhombic borophene, there are four non-zero elastic stiffness constants $C_{11}$, $C_{22}$, $C_{66}$ and $C_{12}$, and the stress-strain relationship is obtained from Hooke's law under plane-stress condition\cite{55}.

\begin{equation}
  \left(\begin{array}{c}
    \sigma_{xx}\\\sigma_{yy}\\\sigma_{xy}
  \end{array}\right)
  =\left(\begin{array}{ccc}
    C_{11}&C_{12}&0 \\ C_{21}&C_{22}&0 \\ 0&0&C_{66}
  \end{array}\right)
  \left(\begin{array}{c}
    \varepsilon_{xx}\\\varepsilon_{yy}\\2\varepsilon_{xy}
  \end{array}\right)
\end{equation}
Where $C_{ij}(i,j=1,2,6)$ is the in-plane stiffness tensor and is equal to the second partial derivative of strain energy $E_\mathrm s$ as a function of strain $\bm{\varepsilon}$ in the range $-2\% < \bm{\varepsilon} <2\%$ with an increment of 0.5\%, based on the following formula\cite{56}
\begin{equation}
  E_\mathrm S = \frac{1}{2}C_{11}\varepsilon_{xx}^2+\frac{1}{2}C_{22}\varepsilon_{yy}^2+C_{12}\varepsilon_{xx}\varepsilon_{yy}+2C_{66}\varepsilon_{xy}^2.
\end{equation}
The engineering strain is defined as $\varepsilon = (L-L_0)/L_0$, where $L$ and $L_0$ are the lattice constants of the strained and unstrained structures, respectively. Here, we get the elastic constants $C_{ij}$ using the VASPKIT code\cite{57}. Then, the Young's modulus $E$ and Poisson's ratio $\nu$ can be derived as\cite{58}
\begin{align}
  E_x=\frac{C_{11}C_{22}-C_{12}C_{21}}{C_{22}}&, E_y=\frac{C_{11}C_{22}-C_{12}C_{21}}{C_{11}},\\
  \nu_{xy}=\frac{C_{21}}{C_{22}}&,\quad \nu_{yx}=\frac{C_{12}}{C_{11}}.
\end{align}

\section{Structural and electronic properties}
\subsection{Chemical bonding nature of borophene}

\begin{figure*}[htb]
	\includegraphics[width=\textwidth]{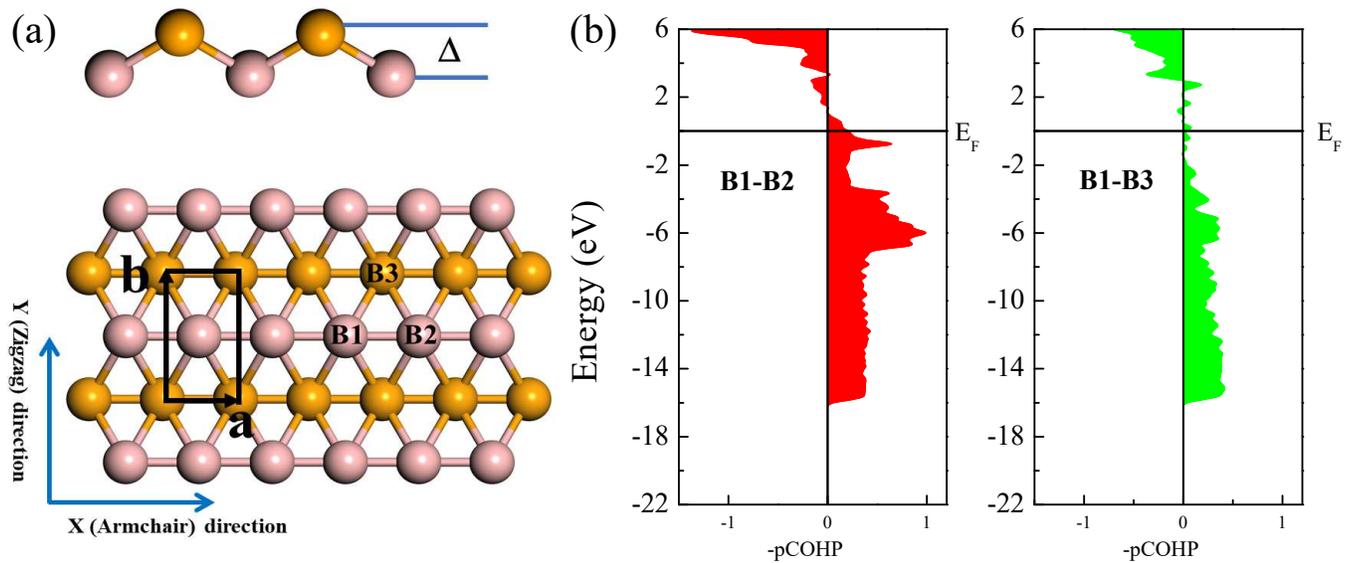}
	\caption{\label{fig1}(a) Side and top views of monolayer borophene. The atoms occupying the A and B sites are on different planes, separated by a distance $\varDelta = 0.941 \mbox\AA$. The blacked rectangle indicates a unit cell. The bond B1--B2 is along the armchair direction, and the bond B1--B3 is along the zigzag direction. (b) COHP curves of monolayer borophene containing B1--B2 and B1--B3 interactions.}
\end{figure*}

The ball-stick structures of monolayer buckled borophene are presented in Fig.~\ref{fig1}(a). Our benchmark calculations on monolayer borophene give lattice constants of $a = 1.613~\mbox\AA$ and $b = 2.880~\mbox\AA$, in good agreement with previous values\cite{13,14,15,26,43}. There is a buckling along $\bm b$ direction with height $\varDelta = 0.941~\mbox\AA$, while no corrugations are observed along $\bm a$ direction. To analyze the nature of the chemical bonding in borophene, Fig.~\ref{fig1}(b) shows our bonding analysis based on -pCOHP. The Fermi level is dominated by B1--B2 interactions. The amount of occupied bonding B--B between B1--B2 atoms is larger than those between B1--B3 atoms. It clearly shows that B1--B2 and B1--B3 bonding interactions consist with the similar interatomic B1--B2 (1.613~$\mbox\AA$) and B1--B3 (1.855~$\mbox\AA$) lengths. A delocalization of electrons over B1 and B3 atoms is through the formation of multi-center bonds along the zigzag direction. Such kind of resonant bonding can also explain the slightly longer bond length compared to classical B--B single bond (1.686~$\mbox\AA$)\cite{44}. On the other hand, B1 and B2 atoms form strong $\sigma$ bond along the armchair direction.

\subsection{Geometry and stability of layered borophene}
\begin{figure*}[htb]
  \includegraphics[width=\textwidth]{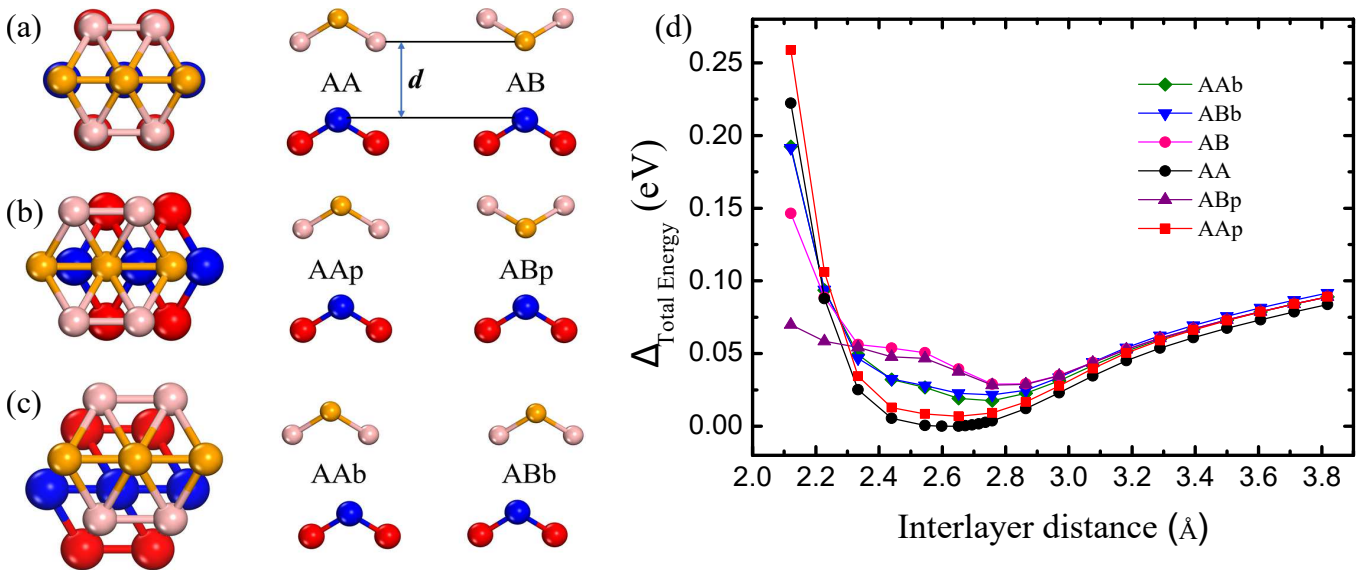}
  \caption{\label{fig2}Six high-symmetry configurations of bilayer borophene are (a) AA and AB with the same top view, (b) AAp and ABp, (c) AAb and ABb. (d) Total energy of these six considered stacking modes as a function of the interlayer distance between two nearest boron atoms of the top and bottom layer along $\bm z$ direction. The interlayer distance varies around the equilibrium distance of each configuration and all atoms are relaxed with this constraint. The lowest calculated total energy (in this case, the total energy associated with the AA configuration) was set to zero, and the others were calculated with respect to this one. }
\end{figure*}

Since the structures of bilayer borophene are rather complicated, we only consider six stacking configurations (AA, AB, AAp, ABp, AAb and ABb) with high symmetry in this work, as shown in Fig.~\ref{fig2}(a)--(c). There are three kinds of top views of these six stacking modes. For AA- and AB-stacking modes, the top layer is directly stacked on the bottom layer. The AAp (ABp) and AAb (ABb)-stacking modes can be viewed as shifting half of the bond length along either B1--B2 or B1--B3 bond direction. Two kinds of side views, coming from the fact that the bottom layer could have the same buckling order as the top or the opposite, have been discussed. All sublayers are initially separated by a distance of 3.0~$\mbox\AA$, and we use the optimized lattice constant of the monolayer borophene as the initial lattice constants for bilayer structures. All structures are totally free to relax (positions of the atoms and the lattice constants).

The lattice parameters of the investigated structures are the same of $a = 1.610~\mbox\AA$ and $b = 2.898~\mbox\AA$, with the same buckling height of 0.941~$\mbox\AA$. The most notable difference among the six stacking modes is the interlayer distance between the top and bottom layers, varying from 3.072~$\mbox\AA$ in the AA-stacking to 3.391~$\mbox\AA$ in the AB-stacking. It can be seen that the energy of the system highly depends on its interlayer distance. To better understand this dependence and to ensure those stacks obtained from the free relaxions reach minimum energy of the system rather than a local minimum, we study the evolution of the total energy as a function of the interlayer distance for each system in Fig.~\ref{fig2}(d). First, these six configurations have an energy minimum without phase transitions. Because different stacking order leads to different $\pi$--$\pi$ interaction distance between delocalized states and thereby different interaction strength, the order of stability of the considered structures is as follows: $\mathrm{AA > AAp > AAb > ABb > AB > ABp}$. AA-stacking structure is found to be the most energetically preferred configuration with the smallest interlayer distance and corresponding strongest interlayer interaction. It is very different from other 2D materials whose preferred configurations are in AB-stacking, such as graphene\cite{45}, silicene\cite{46}, black phosphorene\cite{47}, and hexagonal TMDCs \cite{48}. Furthermore, the corresponding interlayer distance 3.072~$\mbox\AA$ is slightly smaller than the value (3.214~$\mbox\AA$) of AB-stacked phosphorus\cite{47}, which has also a buckled monolayer structure. Based on the AA-stacking bilayer borophene, we take the stacking sequence of AAA and AAAA into trilayer and four-layer borophene structures, respectively.

\subsection{Electronic properties of layered borophene}

\begin{figure*}[htb]
  \includegraphics[width=\textwidth]{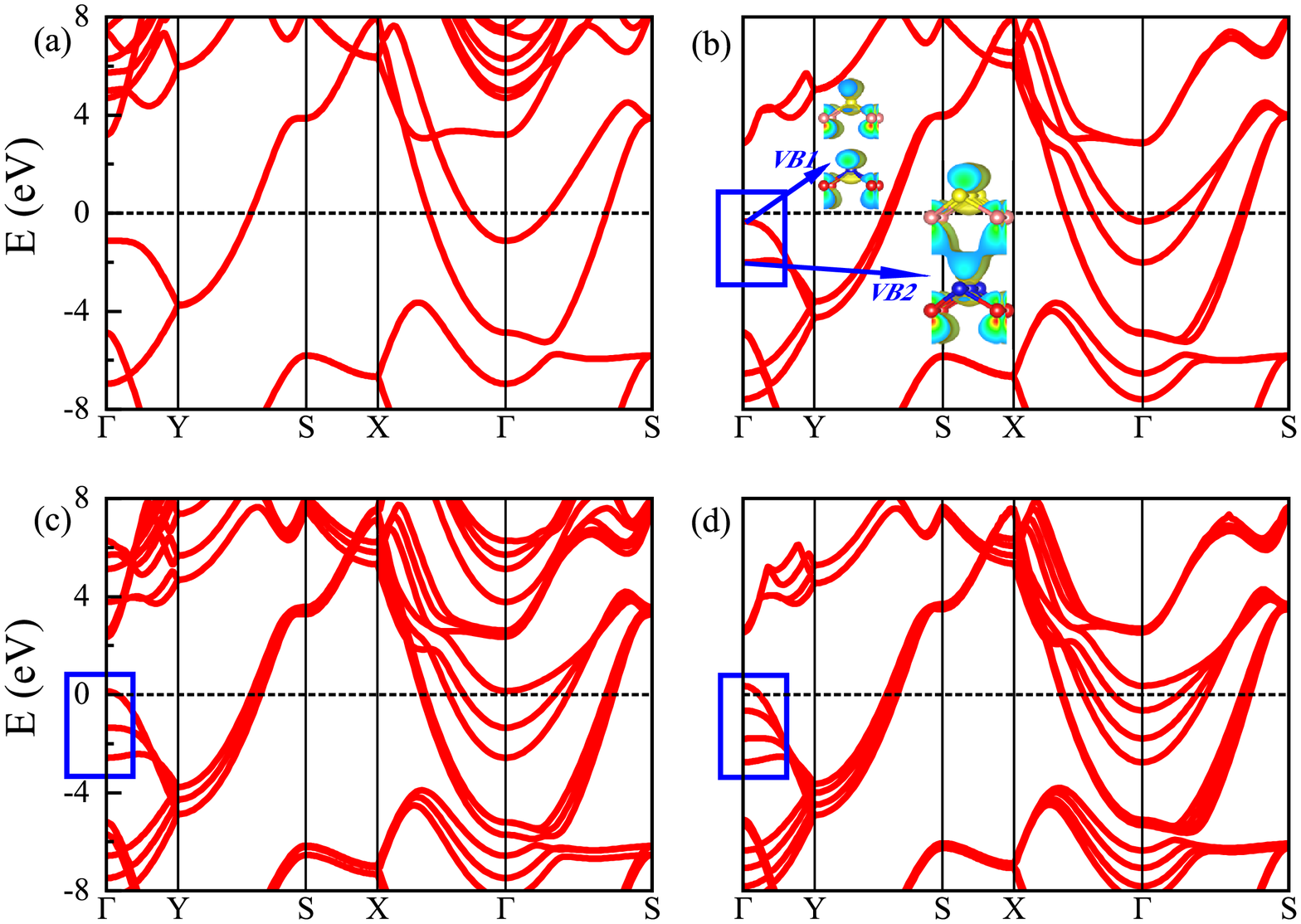}
  \caption{\label{fig3}(a)--(d) Band structures of monolayer, bilayer, trilayer, and four-layer borophene, respectively. The Fermi level is set to zero. Blue rectangles show the band splitting near the Fermi level.}
\end{figure*}

The electronic band structures of monolayer and AA-stacking bilayer, trilayer, and four-layer borophene are shown in Fig.~\ref{fig3}. Compared with the case of monolayer borophene\cite{15}, the Fermi level is crossed by more bands because of the band splitting. Hence, the robust metal feature is retained with the increasing of the layer number. The local band gaps resulting from the buckling along $\Gamma$--Y and S--X directions still exist. The band gap along $\Gamma$--Y direction decreases from 3.263~eV (bilayer) to 2.117~eV (four-layer), as a result of the increasing band splitting with the increasing layer number. Because there is no splitting along S--X direction, the band gap along this high-symmetry line barely depends on the layer number. Therefore, layered borophene behaves anisotropic in electronic properties resulting from the anisotropic atomic structure, and the electrical conductivity is expected to be confined along the uncorrugated armchair direction.

Compared to the band structure of monolayer borophene, the addition of layers results in the band splitting around the $\Gamma$ point in the band structures of layered borophene. The values of band splitting at the $\Gamma$ point for bilayer, trilayer, and four-layer borophene are 1.701, 2.716 and 3.113~eV, respectively. The value of bilayer borophene band splitting is much larger than that (0.5--1~eV) of bilayer MoS$_2$\cite{48} and black phosphorus\cite{47}. This indicates that the interlayer interaction in metallic layered borophene is much stronger than those in other semiconducting 2D materials. To understand the interlayer interaction contribution, we plot the isosurfaces of the charge density corresponding to the VB1 and VB2 of bilayer borophene as insets of Fig.~\ref{fig3}(b), respectively. According to the spatial distribution of the charge density, we can recognize the antibonding and bonding characteristics of the VB1 and VB2 states, which come from the hybridization between the electronic structures of these two sublayers. The bonding characteristics of the VB2 at $\Gamma$ point shows clearly a large overlap of the wave functions from the top and bottom layers, confirming the strong interlayer interaction in layered borophene.

\section{Mechanical properties}

\subsection{Ideal strength and critical strain of layered borophene}

Starting with the optimized borophene structures, tensile strain is applied in either uniaxial (armchair or zigzag) or biaxial direction to explore the ideal tensile strength and the critical strain (the strain at which ideal strength reaches). With each uniaxial strain applied, the lattice constant along the transverse direction and boron atoms are fully relaxed. For biaxial strain, equibiaxial tension is applied and boron atoms in the unit cell are fully relaxed. We calculate the stress-strain relation of 2D layered borophene systems using the method described in the 2D black phosphorene\cite{49}. In a 2D system, the stress is expressed by multiplying the Caucy stresses and $Z/n$ to obtain the equivalent stress, where $Z$ is the thickness of unit cell along the vacuum direction and $n$ is the layer number of the system. To validate our calculations, we compute the mechanical properties of monolayer borophene, such as the elastic stiffness constants and Poisson's ratios shown in Table~\ref{tab1}, which are consistent with previous values\cite{8,43}.

\begin{figure*}[htb]
  \includegraphics[width=\textwidth]{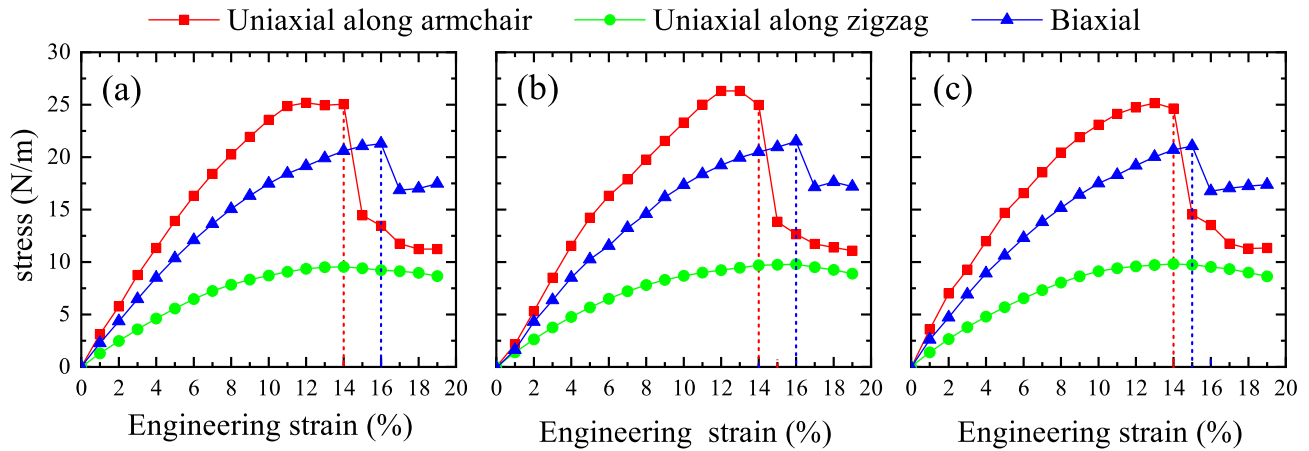}
  \caption{\label{fig4}The stress-strain relations for (a) bilayer (b) trilayer (c) four-layer borophene. The critical strains are 14\% (along the armchair direction) and 15--16\% (along the biaxial direction).}
\end{figure*}

Figs.~\ref{fig4}(a)--(c) present our calculated stain-stress relations. The stress-strain behaviors of layered borophene become nonlinear as the applied strain increases, similar to the case of monolayer structure\cite{43}. From monolayer to layered borophene, the ideal strength along the armchair direction slightly increases from 24.0~N/m (monolayer)\cite{43} to 25.2--26.3~N/m (few-layer). This suggests that the outstanding large tensile strength of borophene is enhanced in layered structure, which is crucial for the mechanical application of few-layer borophene. This enhancement can be explained by the change of the $\sigma$ bond. The $\sigma$ bond length in layered structure is 1.610~$\mbox\AA$, which is shorter than that of monolayer (1.613~$\mbox\AA$). On the other hand, the ideal strengths of layered borophene are 9.5--9.8~N/m along the zigzag direction, smaller than that of monolayer (12.4~N/m). The decrease originates from the enhancement of multi-center bonds by reducing the corresponding bond lengths from monolayer (1.855~$\mbox\AA$)\cite{43} to multilayer (1.836~$\mbox\AA$). For the biaxial tension case, the curve has a maximum value of 21.0~N/m, larger than that of monolayer (19.2~N/m)\cite{43}. The ideal strength of borophene is smaller than those of graphene (36.74--40.41~N/m), but larger than those of silicene (5.26--7.59~N/m), MoS$_2$ (9.59--14.75~N/m), and black phosphorene (4.44--9.99~N/m)\cite{43}.

Unlike the increasing tensile strength along the armchair direction and decreasing strength along the zigzag direction from mono to layered borophene, the critical strain is always increasing in all engineered directions with the increasing number of layers, similar to the trend of black phosphorene\cite{49}. For example, the critical strains are 14\% (armchair), 15--16\% (zigzag), and 14--16\% (biaxial) for few-layer borophene, which are larger than those corresponding critical strains (10\% armchair, 12\% zigzag, 13\% biaxial) in monolayer structure\cite{43}. The increase of critical strains means that the mechanical flexibility of borophene is enhanced from monolayer to multilayer. We note that the increasing critical strains for few-layer borophene are still smaller than those of other 2D materials, such as graphene (19--27\%), black phosphorene (27--33\%), and MoS$_2$ (18--26\%)\cite{43}. To summarize, few-layer borophene exhibit strong anisotropic responses for these three types of applied strains from the stress-strain curves.

\subsection{Buckling height of layered borophene}

\begin{figure*}[htb]
  \includegraphics[width=\textwidth]{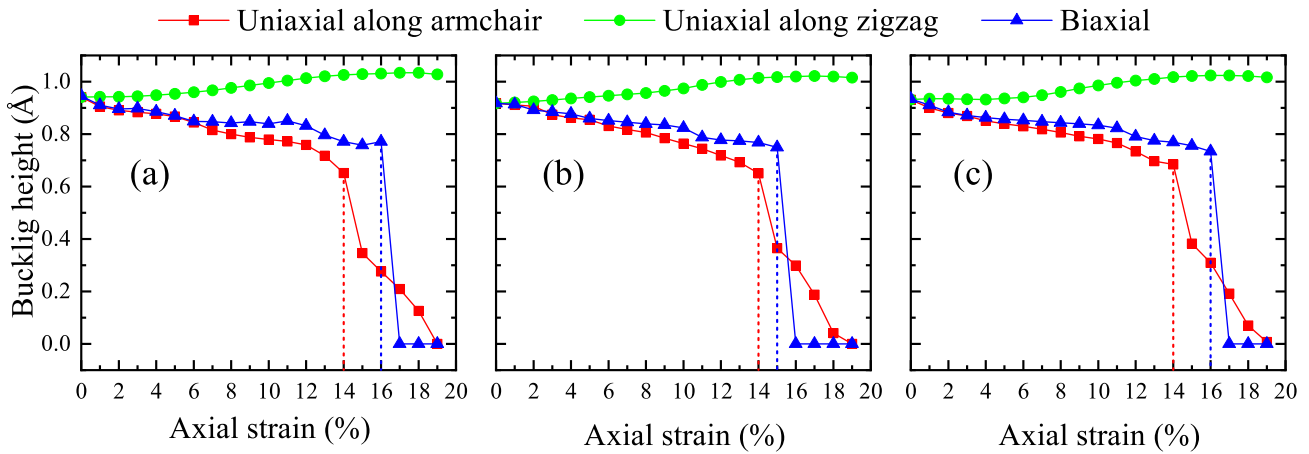}
  \caption{\label{fig5}The calculated dependence of buckling heights of (a) bilayer (b) trilayer (c) four-layer borophene under three types of tension. The buckling heights decrease sharply at the critical strain (14\%) point, and drop to zero at the uniaxial along $\bm a$ of 20\%.}
\end{figure*}

Buckling height is a critical parameter for buckled 2D materials, obviously different from other flat 2D systems. We therefore investigate the dependences of buckling height of layered borophene on three types of applied tension in Fig.~\ref{fig5}. From monolayer to multilayer, the trends of buckling height dependent on tension are nearly the same, showing highly anisotropic and non-monotonic. The buckling heights decrease sharply at the strain of 14\% along the armchair direction, and 15--16\% along the biaxial direction, exactly corresponding to the critical strains of few-layer systems. The layered borophene turns into a graphene-like planar structure instead of the original buckling structure when the strain approaches 19\% (16--17\%) along the armchair (biaxial) direction. Such turning means that the original borophene structure becomes unstable and is destroyed because of the phonon instability. On the contrary, if tension is applied along the zigzag direction, the buckling height increases monotonically with the increasing strain. It means that the out-of-plane Poisson's ratios are negative for layered borophene, similar to monolayer borophene\cite{43}. This is because B1--B3 bonding along the zigzag direction decreases with increasing strain along this direction. The anisotropic out-of-plane Poisson's ratios in few-layer borophene are different from other flat 2D isotropic materials, for example, the layered graphene, $h$-BN, and MoS$_2$ have negative, near zero, and positive out-of-plane Poisson's ratios, respectively\cite{51}.

\subsection{Mechanical constants of layered borophene}

\begin{table*}[htb]
  \caption{\label{tab1}The calculated elastic stiffness constants, Young's modulus, and Poisson's ratios for monolayer and layered borophene. There are four non-zero elastic constants for 2D borophene, because of their orthogonal primitive cell. Here, we get the elastic constants $C_{11}$, $C_{22}$, $C_{66}$ and $C_{12}$ by fitting the data of elastic strain energy $E_{\mathrm s}$ ($\bm{\varepsilon}$) as a function of $\bm{\varepsilon}$ in the strain range $-2\% \leq \bm{\varepsilon} \leq 2\%$ with an increment of 0.5\%. The Young's modulus and Poisson's ratios can be derived from the elastic constants. Poisson's ratio is defined by the ratio of the strain in the vertical direction to that of the applied direction. The calculated values of monolayer structure are in good agreement with previous theoretical results.}
  \begin{ruledtabular}
  \begin{tabular}{lcccrccrr}
    \multirow{3}{*}{System}&\multicolumn{4}{c}{Elastic stiffness constants}&\multicolumn{2}{c}{Young's modulus}&\multicolumn{2}{c}{\multirow{2}{*}{Poisson's ratio}}\\[-1ex]
    &\multicolumn{4}{c}{(GPa$\cdot$nm)}&\multicolumn{2}{c}{(GPa$\cdot$nm)}&\\
    \cline{2-5}\cline{6-7}\cline{8-9}
    &$C_{11}$&$C_{22}$&$C_{66}$&$C_{12}$&$E[100]$&$E[010]$&$\nu$[100]&$\nu$[010]\\
    \hline
    1-layer&396.6&158.4&86.5&$-$3.47&397&158&$-$0.022&$-$0.009\\
    1-layer\cite{59}&377.0&161.0&84.0&1.00&377&162&0.005&0.002\\
    1-layer\cite{59}&405.0&172.0&96.0&$-$1.00&405&172&$-$0.006&$-$0.003\\
    1-layer\cite{8}&398.0&170.0&94.0&$-$7.00&398&170&$-$0.040&$-$0.020\\
    2-layer&380.0&143.8&75.2&7.44&380&144&0.052&0.020\\
    3-layer&361.1&141.5&72.5&11.29&360&141&0.080&0.031\\
    4-layer&337.9&136.0&70.8&12.37&338&136&0.091&0.037\\
  \end{tabular}
  \end{ruledtabular}
\end{table*}

In addition to the stress-strain curves and buckling height dependence, we also calculate elastic constants, Young's modulus, and summarize them together with Poisson's ratios in Table~\ref{tab1}. Due to the anisotropy of the borophene structure, the elastic constants, Young's modulus, and Poisson's ratios have different values along the zigzag and the armchair directions. From monolayer to few-layer borophene, the Young's modulus are decreasing from 396.6~GPa$\cdot$nm (armchair) and 158.4~GPa$\cdot$nm (zigzag) to 337.9~GPa$\cdot$nm (armchair) and 136.0~GPa$\cdot$nm (zigzag). This decreasing trend also appears in the buckled black phosphorus\cite{49}. For both monolayer and few-layer borophene, the Young's modulus along the armchair direction are about 2.5 times larger than their counterparts along the zigzag direction, indicating that it is more difficult to apply strain along the armchair direction. One may notice that the Young's modulus of four-layer borophene along the armchair direction is still very large. This is because the interlayer interactions have negligible influence to the strong $\sigma$ bond along the armchair direction. The large Young's modulus along the armchair direction suggest that few-layer borophene, similar to monolayer borophene, demonstrates super-hardness compared to other 2D materials. This makes borophene a great candidate for practical large-magnitude-strain engineering.

\begin{figure*}[htb]
  \includegraphics[width=\textwidth]{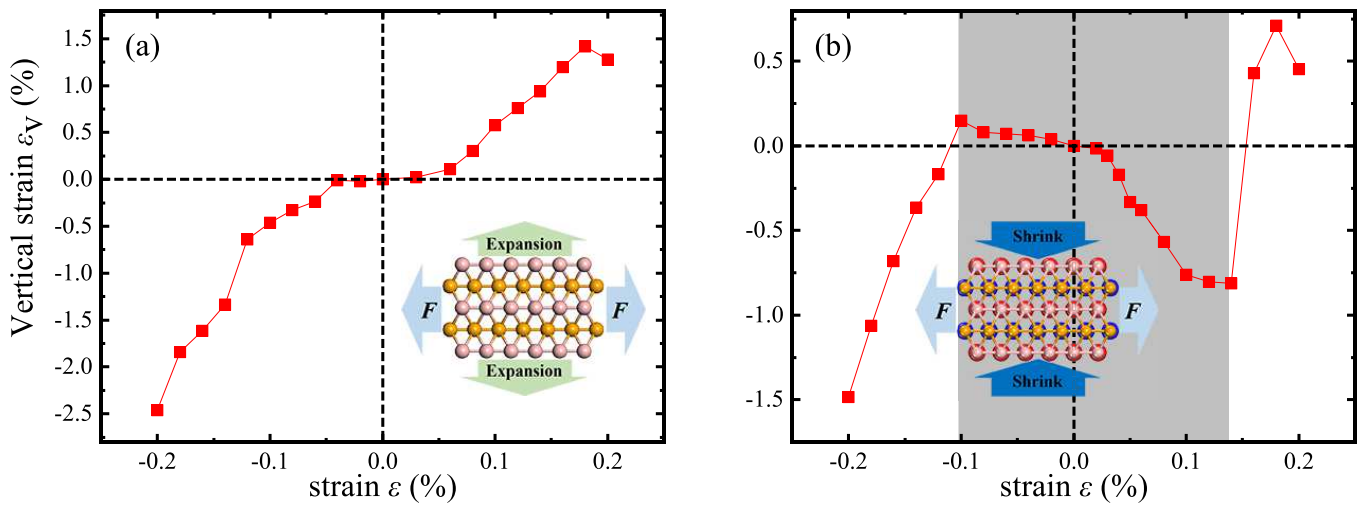}
  \caption{\label{fig6}The vertical strain $\bm{\varepsilon}_\mathrm V$ versus applied strain $\bm{\varepsilon}$ along the armchair direction for monolayer (a) and bilayer (b) borophene, respectively. The positive (negative) $\bm{\varepsilon}$ means a tensile (compressive) strain. The vdW interaction is in principle inverse proportion to the layer distance.}
\end{figure*}

Poisson's ratios measure the fundamental mechanical responses of solid against external loads. The out-of-plane Poisson's ratio, which is related to the change of the buckling height, has been discussed in previous section. We will thus focus on the following in-plane Poisson's ratios in few-layer borophene. For an applied strain $\bm{\varepsilon}$ along the armchair direction in the monolayer, the responding strain $\bm{\varepsilon}_\mathrm V$ occurs along the zigzag direction as shown in Fig.~\ref{fig6}(a). When a strain ($\bm{\varepsilon}$) is applied along the armchair direction, the $\bm{\varepsilon}_\mathrm V$ roughly increases with the increasing strain. That is, the larger the lattice constant $\bm a$ is, the larger the lattice constant $\bm b$ is. If a strain is applied along the zigzag direction, the $\bm{\varepsilon}_\mathrm V$ tends to increase with the increasing strain in the range from -16\% to 10\%. This range is within the critical strains along the zigzag direction for monolayer borophene. Monolayer borophene therefore shows negative in-plane Poisson's ratio of -0.022 along [100] and -0.009 along [010] direction, confirming the anisotropic mechanical properties and in good agreement with previous results\cite{8}.

For few-layer borophene, it is surprising to find that the negative in-plane Poisson's ratio in monolayer borophene dramatically changes into positive, for example, 0.052 along [100] and 0.020 along [010] directions in bilayer borophene, as shown in Table~\ref{tab1}. We should emphasise that this observation is totally different from other 2D materials, even for puckered black phosphorus and arsenene in which the negative Poisson's ratios are preserved from monolayer to multilayer\cite{52,53}. In order to check the reliability of the positive Poisson's ratio in few-layer borophene, we show in Fig.~\ref{fig6}(b) the responding strain $\bm{\varepsilon}_\mathrm V$ for applied strain $\bm{\varepsilon}$ for bilayer borophene as an example, intuitively reflecting the Poisson's ratio. Very different from the case of monolayer borophene, the vertical strain $\bm{\varepsilon}_\mathrm V$ is negative (positive) when the engineered strain is positive (negative) and smaller (larger) than 14\% (-10\%) in bilayer borophene, indicating a positive Poisson's ratio along the armchair direction. For the applied strain along the zigzag direction, the curve of $\bm{\varepsilon}_\mathrm V$ versus $\bm{\varepsilon}$ is similar to the case of armchair direction, indicating also a positive Poisson's ratio. We note that the engineered strains considered here are within the range of the corresponding critical strains, and these results confirm the positive Poisson's ratios shown in Table~\ref{tab1}.

\begin{figure}[htb]
  \includegraphics[width=0.49\textwidth]{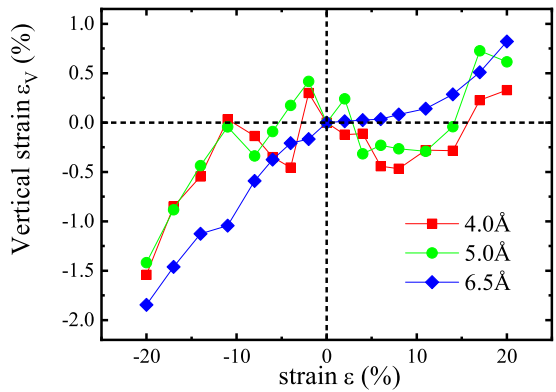}
  \caption{\label{fig7}The vertical strain $\bm{\varepsilon}_\mathrm V$ versus applied strain $\bm{\varepsilon}$ along the armchair direction for varying interlayer distance 4.0 $\mbox\AA$, 5.0 $\mbox\AA$, 6.3 $\mbox\AA$ for bilayer borophene, respectively.}
\end{figure}

The dramatic change of the in-plane Poisson's ratio, i.e., from negative in the monolayer to positive in the multilayer, does not appear in other 2D materials. In fact, this novel phenomenon is a direct consequence of the very strong interlayer vdW interactions appeared in layered borophene, as we discussed in previous sections (see the strong interlayer bonding states shown in Fig.~\ref{fig3}(b)). The interlayer interactions in 2D materials are in general much weaker comparing to the intralayer interactions, even in puckered atomic structures such as black phosphorus and arsenene. As a further check of the influence of the vdW interaction, we investigate the change of Poisson's ratio by varying the interlayer distance in few-layer borophene. Using bilayer borophene as an example, we analyse the curve of vertical strain $\bm{\varepsilon}_\mathrm V$ for applied $\bm{\varepsilon}$ along the armchair direction, as displayed in Fig.~\ref{fig7}. It is clear that with different interlayer distance, the curves of vertical strain $\bm{\varepsilon}_\mathrm V$ versus $\bm{\varepsilon}$ become very different. When the interlayer distance approaches 6.3 $\mbox\AA$, the curve becomes similar to that of monolayer borophene, with the negative Poisson's ratio recovered. This confirms that the vdW interlayer interactions are much stronger in metallic borophene than that in semiconducting 2D materials (such as flat TDMCs and puckered black phosphorus), resulting in a shorter interlayer distance and therefore intensively altering the mechanical properties of few-layer borophene. Furthermore, as discussed in Ref.~\onlinecite{51,54}, the energy of the interlayer vdW interactions for 2D thick metals is proportional to $d^{-2}$ (where $d$ is the interlayer distance), while the asymptotic vdW energy of parallel structures is proportional to $d^{-4}$ for 2D insulators. Thus, the decay speed of vdW interactions in few-layer metallic borophene is significantly slower than that in semiconducting black phosphorus and arsenene.

\section{Conclusion}
In conclusion, we have studied electronic and mechanical properties of few-layer borophene based on the buckled 2-P$mmm$ monolayer structure as synthesized by Mannix \emph{et al} in Ref.~\onlinecite{8}. We find that the AA-stacking mode is the most stable one among the six high-symmetry stacking configurations for bilayer structures. From mono to layered borophene, the robust anisotropic metallic features are maintained, with large energy splitting at $\Gamma$ point ($\sim$1.5 eV), confirming strong interlayer vdW interactions. Since the layered structures can withstand larger critical strains than that in monolayer, layered borophene exhibit more flexibility than monolayer one. Because of the preserved multi-center bonds along the zigzag direction, the out-of-plane negative Poisson's ratios are preserved. In contrast, the in-plane negative Poisson's ratios in the monolayer become positive in layered borophene. This novel phenomenon is a direct consequence of the very strong vdW interlayer interactions, and the negative Poisson's ratios could recover if the interlayer distance is increased to 6.3 $\mbox\AA$ artificially. The dramatic change of the in-plane Poisson's ratio from monolayer to multilayer does not appear in other 2D materials, even in puckered black phosphorus and arsenene. We hope that our theoretical results will inspire considerable experimental enthusiasm of few-layer borophene, especially for potential applications in novel electronic and mechanical devices.

\begin{acknowledgments}
We acknowledges the financial support from Thousand Young Talent Plan (China), and also thank the supercomputing system in the Supercomputing Center of Wuhan University for our numerical calculations.
\end{acknowledgments}

\bibliographystyle{apsrev4-1}
\bibliography{borophene}

\end{document}